\documentclass[doublecol,linenumbers]{epl2} 
\usepackage{graphicx}
\usepackage{dcolumn}
\usepackage{bm}
\usepackage{graphicx,amssymb,amsmath,amsthm,amsfonts,epsfig}

\title{ Logarithmic Corrections to the Black Hole Entropy Product of ${\cal H}^{\pm}$ via Cardy Formula }

\author{Parthapratim Pradhan\footnote{pppradhan77@gmail.com}\inst{1}}

\institute{                    
  \inst{1} Department of Physics, Vivekananda Satavarshiki Mahavidyalaya,
West Midnapur, West Bengal 721513, India \\
}

\abstract{
We compute the logarithmic corrections to  black hole (BH) entropy product of ${\cal H}^{\pm}$ \footnote{ ${\cal H}^{+}$  
and ${\cal H}^{-}$ denote outer (event) horizon and inner (Cauchy) horizons} by using \emph{Cardy prescription}. We particularly 
apply this formula for \emph{BTZ BH}. We speculate that the logarithmic corrections to entropy product of 
${\cal H}^{\pm}$ when computed \emph{via Cardy formula}  the product should be neither \emph{mass-independent (universal)} 
nor be \emph{quantized}.}

\begin{document}

\maketitle

\section{Introduction}
In 1973, a remarkable result was predicted by Bekenstein \cite{bk73} and Hawking \cite{bcw73} in the 
first history of science that the entropy of a dark star (so called BH) is proportional to the 
``geometric quantity'', so called the area of event horizon (EH). This immediately suggests that the outer BH entropy
\footnote{We know that when $\hslash= 0$, Quantum mechanics reduces to classical mechanics. If we take this limit 
in ${\cal S}_{+}$ we have divergence value of the outer entropy. Therefore we can say that there is no 
classical BH entropy. This ${\cal S}_{+}$ is a purely \emph{quantum BH entropy} for ${\cal H}^{+}$. Similarly, we 
can suggest that ${\cal S}_{-}$ is a purely \emph{quantum inner BH entropy}.} is given by 
\begin{eqnarray}
{\cal S}_{+} &=& \frac{k_{B}c^3}{\hslash}\frac{{\cal A}_{+}}{4G} ~\label{cd1}
\end{eqnarray}
where $k_{B}$ is the Boltzman constant from statistical mechanics, $c$ is the
speed of light in free space come from special theory of relativity, $\hslash$ is called reduced Planck 
constant and comes from quantum mechanics, ${\cal A}_{+}$ is the area of ${\cal H}^{+}$ come from 
purely geometry of the spacetime and $G$ is a universal constant that comes from gravity. 

If a BH has another horizon so called inner horizon or Cauchy horizon (${\cal H}^{-}$) there must 
exists \emph{inner BH entropy} which can be defined as
\begin{eqnarray}
{\cal S}_{-} &=& \frac{k_{B}c^3}{\hslash}\frac{{\cal A}_{-}}{4G} ~\label{cd2}
\end{eqnarray}
where ${\cal A}_{-}$ is the area of ${\cal H}^{-}$ also come from inner geometry.

Now the product of outer BH entropy and inner BH entropy should read 
\begin{eqnarray}
{\cal S}_{+} {\cal S}_{-} &=& \left(\frac{k_{B}c^3}{\hslash G}\right)^2 \frac{{\cal A}_{+}{\cal A}_{-}}{16}  
~\label{cd3}
\end{eqnarray}
which implies that it is proportional to the product of the geometric quantity of  ${\cal H}^{\pm}$. Now if we 
define the fundamental length scale so called Planck length, i.e.,  
\begin{eqnarray}
\ell_{Pl} &=& \sqrt{\frac{G \hslash}{c^3}} ~\label{cd4}
\end{eqnarray}
then the products of BH entropy as 
\begin{eqnarray}
{\cal S}_{+} {\cal S}_{-} &=& \frac{{\cal A}_{+}{\cal A}_{-}}{16 \ell_{Pl}^4}  ~\label{cd5}
\end{eqnarray}
where we have to set $k_{B}=1$.

This area (or entropy) product formula of ${\cal H}^{\pm}$  for wide class of BHs 
\cite{ah09,cgp11,castro12,sd12,mv13,val13,pp14,jh,horava15,grg16,grg1} has been examined so far with out 
taking into account any logarithmic corrections. Without logarithmic corrections the product of area (or entropy) 
of ${\cal H}^{\pm}$ is universal in some cases and it fails to be universal in some cases also. But it is interesting
when we have taken the logarithmic corrections of this product then the  product should always not be universal. 
Our aim is here to derive the logarithmic correction to the BH entropy of ${\cal H}^{\pm}$ and its product 
\emph{via Cardy prescription}
\cite{cardy,cardy1,vafa,ajak,carlip,carlip1,carlip2,carlip3,carlip4,carlip5,carlip6,solo,km,km1,skm,psm,suneeta,kk}.

On the other hand in the framework of String theory for BPS (Bogomol'ni-Prasad-Sommerfield) class of BHs there has been 
a proposal that the product of inner and outer BH entropy is quantized in nature \cite{cgp11} and 
it should be
\begin{eqnarray}
{\cal S}_{+} {\cal S}_{-}  &=& 
\left(2\pi \ell_{Pl}^2\right)^2 N , \,\, N \in {\mathbb{N}} ~.\label{cd6}
\end{eqnarray}
It should be noted that Larsen \cite{finn}  proposed that the BH outer horizon 
as well as inner BH horizon is quantized in the units of Planck. That means the product of inner area (or inner entropy) 
and outer area (or outer entropy) of ${\cal H}^{\pm}$ is quantized in terms of Planck units. 

In the next section, we will calculate the logarithmic corrections to BH entropy product formula by using the
\emph{Cardy formula}.

\section{Logarithmic Corrections to BH Entropy Product Formula via Cardy method}
In order to compute the logarithmic corrections to the density of states of ${\cal H}^{\pm}$, we begin with an arbitrary 2D 
CFT with central charges $c$ by using the Virasoro algebra of ${\cal H}^{\pm}$ \cite{brown,fran,ralph}
$$
\left[L_{m, \pm}, L_{n, \pm} \right] = \left(m-n\right) L_{m+n, \pm}+\frac{c}{12} m \left(m^2-1\right) \delta_{m+n, 0} 
$$
$$
\left[\tilde{L}_{m, \pm}, \tilde{L}_{n, \pm} \right]  = \left(m-n\right) \tilde{L}_{m+n, \pm}+\frac{c}{12} m \left(m^2-1\right)
\delta_{m+n, 0}\\
$$
\begin{eqnarray}
\left[L_{m, \pm}, \tilde{L}_{n, \pm} \right] &=& 0 ~ \label{eq1}
\end{eqnarray}
where the generators $L_{n, \pm}$ and $\tilde{L}_{n, \pm}$ are ``holomorphic'' and ``anti-holomorphic'' diffeomorphisms, 
respectively.

At the same time we can define the partition function of  ${\cal H}^{\pm}$ on the 2-torus  of modulus 
$\tau=\tau_{1}+i\tau_{2}$ is defined to be 
\begin{eqnarray}
{\cal Z}_{\pm} (\tau, \tilde{\tau}) &=& Tr \, e^{2\pi i\tau L_{0, \pm}} e^{-2\pi i \tilde{\tau} \tilde{L}_{0,\pm}} \nonumber\\
&=& \sum \rho_{\pm} \left(\Delta_{\pm}, \tilde{\Delta}_{\pm} \right) e^{2\pi i\tau \Delta_{\pm}} e^{-2\pi i \tilde{\tau} 
\tilde{\Delta}_{\pm}} ~\label{eq2}
\end{eqnarray}
where $\rho_{\pm}$ is the number of states with eigen values $L_{0, \pm}=\Delta_{\pm}$, $\tilde{L}_{0,\pm}=\tilde{\Delta}_{\pm}$.

Now if we can compute the partition function ${\cal Z}_{\pm}$, we can calculate the density of states $\rho_{\pm}$ 
via contour integration. For this we can assume $q=e^{2\pi i\tau}$ and $\tilde{q}=e^{2\pi i\tilde{\tau}}$. Therefore one 
should find the contour integration for density of states of ${\cal H}^{\pm}$
\begin{eqnarray}
\rho_{\pm} \left(\Delta_{\pm}, \tilde{\Delta}_{\pm} \right)  &=& \frac{1}{(2\pi i)^2} \int \frac{dq}{q^{\Delta_{\pm}+1}}
\frac{d\tilde{q}}{\tilde{q}^{\tilde{\Delta}_{\pm}+1}}
{\cal Z}_{\pm} (q, \tilde{q})  ~\label{eq3}
\end{eqnarray}
where the contour integration evaluated from $q=0$ to $\tilde{q}=0$. Actually Cardy \cite{cardy,cardy1} found that the 
partition function of ${\cal H}^{\pm}$ is given by 
\begin{eqnarray}
{\cal Z}_{\pm} (\tau, \tilde{\tau}) &=& \frac{Tr \, e^{2\pi i \left(L_{0, \pm}-\frac{c}{24}\right)\tau} 
e^{-2\pi i \left(\tilde{L}_{0, \pm}-\frac{c}{24}\right)\tilde{\tau}}}{e^{\frac{\pi c}{6}\tau_{2}}} 
~. \label{eq4}
\end{eqnarray}
Interestingly, this quantity is ``modular-invariant''. It is also universal via CFT. Using this result we can evaluate 
the above integral by steepest descent method. Now let $\Delta_{0, \pm}$ be the lowest eigen value of $L_{0, \pm}$ and 
define 
\begin{eqnarray}
\bar{{\cal Z}}_{\pm} (\tau) &=& \sum \rho_{\pm} \left(\Delta_{\pm} \right) e^{2\pi i \left(\Delta_{\pm}-\Delta_{0, \pm}\right)\tau}
\nonumber\\
&=& \rho_{\pm} \left(\Delta_{0, \pm}\right)+ \rho_{\pm} \left(\Delta_{1, \pm}\right) 
e^{2\pi i \left(\Delta_{1, \pm}-\Delta_{0, \pm} \right)\tau} +... ~. \nonumber\\
\label{eq5}
\end{eqnarray}
For convenient, we have omitted the $\tilde{\tau}$ dependence. Then it can easily be shown that 
\begin{eqnarray}
\rho_{\pm} \left(\Delta_{\pm}\right)=  \int e^{\frac{2\pi i}{\tau}\left(\frac{c}{24}-\Delta_{0, \pm} \right)} 
e^{2\pi i\tau \left(\frac{c}{24}-\Delta_{\pm}\right)} 
\bar{{\cal Z}}_{\pm} \left(-\frac{1}{\tau}\right) d\tau ~.\label{eq6}
\end{eqnarray}
For large value of $\tau_{2}$, it can be shown that $\bar{{\cal Z}}_{\pm} \left(-\frac{1}{\tau}\right)$ gives us a constant 
value $\rho_{\pm} \left(\Delta_{0, \pm}\right)$. Therefore the above integral becomes 
\begin{eqnarray}
\rho_{\pm} \left(\Delta_{\pm}\right)  &\approx&  \left(\frac{c}{96 \Delta_{\pm}^3} \right)^{\frac{1}{4}}
e^{2\pi \sqrt{\frac{c\Delta_{\pm}}{6}}} ~.\label{eq7}
\end{eqnarray}
Now one can obtain the exponential part of the Eq. (\ref{eq7}) which is actually the Cardy formula. Now, 
one can  apply this formula for calculating the entropy of ${\cal H}^{\pm}$ for rotating BTZ BH and compared it with the 
result obtained by Strominger  in his work \cite{strom}.

The BH event horizon and Cauchy horizon for rotating BTZ BH \cite{btz92,jetpl} is given by
\begin{equation} 
r_{\pm}= \sqrt{4G_{3}{\cal M} \ell^2\left(1\pm \sqrt{1-\frac{J^2}{{\cal M}^2 \ell^2}} \right)}
 ~\label{eq8}
\end{equation}
where $G_{3}$ is the 3D Newtonian constant. Now it can be easily derived the ADM mass parameter and angular 
momentum parameter 
\begin{equation} 
M = \frac{r_{+}^2+r_{-}^2}{8G_{3}\ell^2}, \,\,\, J=\frac{r_{+}r_{-}}{4G_{3}\ell}  ~\label{eq9}
\end{equation}
where $\ell^2=-\frac{1}{\Lambda}$ and $\Lambda$ is cosmological constant. The central charges derived by Brown and Henneaux 
\cite{brown} using the properties of asymptotic symmetries in 3D with negative cosmological constant which could be determined 
by a pair of Virasoro algebra,  are
\begin{eqnarray}
 c &=& \tilde{c}=\frac{3\ell}{2G_{3}}   ~.\label{eq10}
\end{eqnarray}
The generators of the Brown and Henneaux Virasoro algebras derived in \cite{bana} are 
\begin{eqnarray}
\Delta_{\pm} &=& \frac{\left(r_{\pm}+r_{\mp} \right)^2}{16 G_{3} \ell}, \,\,\, 
\tilde{\Delta}_{\pm} = \frac{\left(r_{\pm}-r_{\mp} \right)^2}{16 G_{3} \ell}~.\label{eq11}
\end{eqnarray}
Using Eq. (\ref{eq10}) and  Eq. (\ref{eq11}), one can compute the exponential part in Eq. (\ref{eq7}) as
\begin{eqnarray}
 2\pi \sqrt{\frac{c \Delta_{\pm}}{6}}+2\pi \sqrt{\frac{\tilde{c} \tilde{\Delta}_{\pm}}{6}} &=& \frac{2\pi r_{\pm}}{4G_{3}}
 ~\label{eq12}
\end{eqnarray}
which gives the standard Bekenstein-Hawking entropy for 3D BH and it was first observed by 
Strominger \cite{strom} for  the ${\cal H}^{+}$ in 1998. We examined here this entropy calculation 
is valid for both ${\cal H}^{\pm}$. 

Using  Eq. (\ref{eq7}), one can easily compute the density of states of ${\cal H}^{\pm}$ as 
\begin{eqnarray}
\rho_{\pm} \left(\Delta_{\pm}, \tilde{\Delta}_{\pm} \right)  &\approx& \frac{8G_{3}\ell^2}{\left(r_{\pm}^2-r_{\mp}^2\right)^
{\frac{3}{2}}} e^{\frac{2\pi r_{\pm}}{4G_{3}}}  ~.\label{eq13}
\end{eqnarray}

Therefore, one should calculate the logarithmic corrections to the entropy of ${\cal H}^{\pm}$
\begin{eqnarray}
{\cal S}_{\pm}  &\sim& \frac{2\pi r_{\pm}}{4G_{3}}-\frac{3}{2} \ln \left|\frac{r_{\pm}^2-r_{\mp}^2}{ G_{3}^2} \right|+const.\\
&=& \frac{2\pi r_{\pm}}{4G_{3}}-\frac{3}{2} \ln \left|\frac{2\pi r_{\pm}}{G_{3}} \right|-
\frac{3}{2} \ln \left|\kappa_{\pm} \ell\right|+const.~ \label{eq14}
\end{eqnarray}
where the surface gravity is defined to be 
\begin{eqnarray}
\kappa_{\pm}  &=& \frac{r_{\pm}^2-r_{\mp}^2} {\ell^2 r_{\pm}}  ~.\label{eq15}
\end{eqnarray}

Therefore the logarithmic terms in Eq. (\ref{eq14}) obtained by Kaul and Majumdar \cite{km1} for ${\cal H}^{+}$ for spherically 
symmetric BH in 4D  have exactly the same form as we have seen from the above calculation. It should be noted that 
this is  also  valid for ${\cal H}^{-}$. Thus one can compute their product and should read off
$$
{\cal S}_{+} {\cal S}_{-}=\frac{\pi^2}{4G_{3}} r_{+}r_{-} 
-\frac{3\pi}{4G_{3}}\left[r_{+}\ln \left|\frac{2\pi r_{-}}{G_{3}} \right|+r_{-}\ln \left|\frac{2\pi r_{+}}{G_{3}} \right|\right]
$$
$$
-\frac{3\pi}{4G_{3}}\left[r_{+}\ln \left| \kappa_{-} \ell\right|+r_{-}\ln \left|\kappa_{+} \ell \right|\right]
$$
$$
+\frac{9}{2}\left[ \ln \left|\frac{2\pi r_{+}}{G_{3}} \right| \ln \left| \kappa_{-} \ell\right|+
\ln \left|\frac{2\pi r_{-}}{G_{3}} \right| \ln \left| \kappa_{+} \ell\right|\right]
$$
\begin{eqnarray}
+\frac{9}{4} \ln \left|\frac{2\pi r_{+}}{G_{3}} \right| \ln \left|\frac{2\pi r_{-}}{G_{3}} \right|
+\frac{9}{4} \ln \left| \kappa_{+} \ell\right| \ln \left| \kappa_{-} \ell\right|+const.  ~.  \label{eq16}
\end{eqnarray}
It follows from the above analysis that without logarithmic correction the product of entropy is always mass-independent
(universal) but the problem is when we have taken into account the logarithmic correction term the product of ${\cal H}^{\pm}$
always dependent on the mass parameter that means it is not universal as well it is not quantized. This is the key result of this 
work.

So far we have examined the logarithmic corrections to the entropy product formula of ${\cal H}^{\pm}$ using Cardy formula 
now we shall calculate  the entropy using slightly different CFT described by the universal Virasoro algebra at the 
horizon \cite{carlip,carlip1,carlip2,carlip3,solo} with central charge 
\begin{eqnarray}
c  &=& \frac{3 {\cal A}_{\pm}\beta_{\pm}}{2\pi G \kappa_{\pm}}  ~\label{eq17}
\end{eqnarray}
and an $L_{0, \pm}$ eigenvalue is
\begin{eqnarray}
\Delta_{\pm}  &=& \frac{{\cal A}_{\pm}\kappa_{\pm}}{16\pi G \beta_{\pm}}  ~.\label{eq18}
\end{eqnarray}
where ${\cal A}_{\pm}$ is the horizon area of ${\cal H}^{\pm}$, $\kappa_{\pm}$ is the surface gravity of ${\cal H}^{\pm}$
and $\beta_{\pm}$ is periodicity of ${\cal H}^{\pm}$.

Now,  reverting back these values in  Eq. (\ref{eq7}), one obtains the density of states of 
${\cal H}^{\pm}$ as 
\begin{eqnarray}
\rho_{\pm} \left(\Delta_{\pm}\right)  &\approx&  \frac{c}{12}
\frac{e^\frac{{\cal A}_{\pm}}{4G}}{\left(\frac{{\cal A}_{\pm}}{8\pi G} \right)^{\frac{3}{2}}} ~.\label{eq19}
\end{eqnarray}
For $c$ to be a universal constant one must need to be choose the value of $\beta_{\pm}$  such that it is independent 
of ${\cal A}_{\pm}$ then one obtains the entropy of ${\cal H}^{\pm}$:
\begin{eqnarray}
{\cal S}_{\pm} &\sim& \frac{{\cal A}_{\pm}}{4G}-\frac{3}{2}\ln\left(\frac{{\cal A}_{\pm}}{4G}\right) +const.+ ... ~\label{eq20}
\end{eqnarray}
where we have set $\ell_{Pl}^2=1$. Interestingly, the above  Eq. (\ref{eq20}) is completely  in 
agreement with the result of Kaul and Majumdar obtain for ${\cal H}^{+}$ only  \cite{km1}. 
We have suggested here this entropy expression is valid for both ${\cal H}^{\pm}$.

To summarize, we computed  the logarithmic corrections to the BH entropy of inner and outer horizons, and their 
product by using the trick of Cardy formula. We have considered particularly rotating BTZ BH and showed when we have taken 
into account the logarithmic corrections to the entropy product it should not quite independent of the ADM mass parameter 
henceforth it should not be quantized.

\bibliographystyle{model1-num-names}

\end{document}